\documentclass[apj]{emulateapj}   
\usepackage{graphicx}
\usepackage{multirow}

\usepackage{amsmath}
\usepackage{natbib}
\usepackage{hyperref}
\usepackage[usenames,dvipsnames]{color}  

\usepackage[normalem]{ulem}

\usepackage{lineno}

\def\msun{{\rm ~M}_{\odot}}
\def\rsun{{\rm ~R}_{\odot}}
\def\zsun{{\rm ~Z}_{\odot}}
\def\gpy{{\rm ~Gpc}^{-3} {\rm ~yr}^{-1}}
\def\kms{{\rm ~km} {\rm ~s}^{-1}}

\newcommand{\mcl}{\ensuremath{M_{\rm cl}}}
\newcommand{\rh}{\ensuremath{r_{\rm h}}}
\newcommand{\nbseven}{{\tt NBODY7}}

\newcommand{\bse}{{\tt BSE}}

\newcommand{\fbin}{\ensuremath{f_{\rm bin}}}
\newcommand{\fobin}{\ensuremath{f_{\rm Obin}}}
\newcommand{\archain}{{\tt ARCHAIN}}

\definecolor{chmagenta}{rgb}{0.54, 0.17, 0.88}

\begin{document}

\title{Black hole---black hole total merger mass and the origin of LIGO/Virgo sources} 

\author{
   Krzysztof Belczynski\altaffilmark{1}, Zoheyr Doctor\altaffilmark{2}, Michael Zevin\altaffilmark{3,4}, 
   Aleksandra Olejak\altaffilmark{1}, Sambaran Banerjee\altaffilmark{4,5}, Debatri 
   Chattopadhyay\altaffilmark{6}
}

\affil{
   $^{1}$ Nicolaus Copernicus Astronomical Center, Polish Academy of Sciences, Bartycka 18, 
          00-716 Warsaw, Poland (chrisbelczynski@gmail.com, aleksandra.olejak@wp.pl)\\
   $^{2}$ Center for Interdisciplinary Exploration and Research in Astrophysics (CIERA) and 
          Department of Physics \& Astronomy, Northwestern University, 1800 Sherman Ave, Evanston, 
          IL 60201, USA (zoheyr.doctor@gmail.com)\\
   $^{3}$ Kavli Institute for Cosmological Physics, The University of Chicago, 5640 South Ellis Avenue, Chicago, Illinois 60637, USA \\
   $^{4}$ Enrico Fermi Institute, The University of Chicago, 933 East 56th Street, Chicago, Illinois 60637, USA
          (michael.j.zevin@gmail.com)\\
   $^{5}$ Helmholtz-Instituts f\"ur Strahlen- und Kernphysik (HISKP), Nussallee 14-16, D-53115 Bonn, 
          Germany (sambaran.banerjee@gmail.com)\\
   $^{6}$ Argelander-Institut f\"ur Astronomie (AIfA), Auf dem H\"ugel 71, D-53121 Bonn, Germany\\ 
   $^{7}$ Gravity Exploration Institute, School of Physics and Astronomy, Cardiff University, 
          Cardiff, CF24 3AA, UK (chattopadhyaydebatri@gmail.com) 
}

\begin{abstract}
The LIGO-Virgo-KAGRA (LVK) Collaboration has reported nearly 100 BH-BH mergers.  
The LVK provides estimates of rates, masses, effective spins, and redshifts for 
these mergers. Yet, the formation channel(s) of the mergers remains uncertain. 
One way to search for a formation site is to contrast properties of detected 
BH-BH mergers with different models of BH-BH merger formation.
Our study is designed to investigate the usefulness of the total BH-BH merger mass
and its evolution with redshift in establishing the origin of gravitational-wave sources.
We find that the average {\em intrinsic} BH-BH total merger mass shows exceptionally 
different behavior for the  models that we adopt for our analysis. In the local universe 
($z=0$) the average merger mass changes from $\overline{M}_{\rm tot,int}\sim25\msun$ for 
CE binary evolution and open clusters formation channels, to $\overline{M}_{\rm tot,int}\sim30\msun$ 
for the stable-RLOF binary channel, to $\overline{M}_{\rm tot,int}\sim45\msun$ for the globular 
cluster channel. These differences are even more  pronounced at larger redshifts. However, 
these differences are diminished when considering LVK O3 detector sensitivity. Comparison 
with LVK O3 data shows that none of our adopted models can match the data despite large 
errors on BH-BH masses and redshifts.
We emphasize that our conclusions are derived from  a small set of $6$ models that are subject 
to numerous known uncertainties. We also note that BH-BH mergers may originate from a mix of 
several channels and that other (than those adopted here) BH-BH formation channels may exist. 
\end{abstract}


\section{Introduction}
\label{sec:introduction}

The latest LIGO-Virgo-KAGRA (LVK) Collaboration catalog, GWTC-3 \citep{GWTC3a}, reports the detection of 
$90$ double compact object mergers, $83$ of which are confidently classified as binary black hole 
mergers. The LVK Collaboration also provided estimates of the intrinsic population properties of these 
mergers such as their distributions of rates, masses, and effective spins. The merger rate density of BH-BH 
systems at a fiducial redshift of $z=0.2$ is $17.3-45\gpy$ and the rate increases with redshift \citep{GWTC3b}. 
Masses of BH-BH mergers show complex multi-feature behavior \citep{TiwariStructure,Edelman,GWTC3b}. BH-BH 
merger effective spins are mostly positive and small ($0<\chi_{\rm eff}<0.3$) with $\sim 10-30\%$ contribution 
of mergers with high positive ($\chi_{\rm eff}>0.3$) effective spins \citep{Roulet2021,Abbott2021a,
Abbott2021b}. 

Despite all the available information and rapidly increasing number of detected BH-BH mergers, their 
origin and specific formation mechanism is not known. Several formation channels are under close 
investigation, each corresponding to a specific formation site of BH-BH mergers and specific evolutionary 
processes involved in the evolution of BH-BH progenitors. Evolution of binary systems in isolation 
(galactic fields) can provide ample resources for BH-BH merger formation via several sub-channels; 
evolution that involves a common envelope phase (CE, e.g., \citealt{Belczynski2016b,Giacobbo2018,
Kinugawa2020,Garcia2021}), evolution that only requires binary interactions through stable Roche-lobe 
overflow (RLOF, e.g., \citealt{Heuvel2017,Neijssel2019,Bavera2021,Olejak2021a}), and chemically 
homogeneous evolution that requires rapidly rotating stars (e.g., \citealt{Marchant2016,deMink2016,
Mandel2016a}). Evolution of objects in dense stellar systems may also provide satisfactory conditions 
for dynamical formation of BH-BH mergers; in young open clusters (e.g., \cite{Banerjee2018,DiCarlo2020,
Kremer2020b,Rastello2021,Chattopadhyay2022}), in dense globular cluster (e.g., \citealt{Rodriguez2016a,
Askar2017,Antonini2020}), and in massive nuclear clusters (e.g., \citealt{ArcaSedda2017,ArcaSedda2020,
Fragione2021,Mapelli2021}). There are also formation channels that bridge the two above major classes of 
isolated and dynamical BH-BH formation. Multiple systems, in particular triples and quadruples, are also 
estimated to produce BH-BH mergers through combination of typical stellar evolution assisted by dynamical 
interactions of stars (e.g., ~\citealt{Antonini2017b,Safarzadeh2020}). Alternatively, gas assisted evolution 
in AGN disks may provide a way to form merging BH-BH systems (e.g., \citealt{Samsing2020,Ford2021}). 

Identifying the actual BH-BH merger formation channel or channels is hindered by numerous modeling 
uncertainties \citep{ArcaSedda2019,Belczynski2022,Broekgaarden2022,GallegosGarcia2021}. For example, 
merger rates for double compact objects vary by orders of magnitude for each formation channel (e.g., 
\citealt{Mandel2021}), while predicted mass distributions are extremely sensitive to various model 
assumptions (e.g., \citealt{Olejak2021a}). Ideally, a given model should fit all physical properties 
(rates, masses, spins) for all three double compact object types (BH-BH/BH-NS/NS-NS mergers; NS: neutron 
star). Before this is accomplished, we need to build understanding of how various model parameters change 
properties of double compact objects within various formation channels. Numerous studies were already 
dedicated to merger rates, NS/BH spins, or compact object masses (e.g., \cite{Timmes1996,Lipunov1997,
Belczynski2002,Voss2003,Podsiadlowski2003,Zampieri2009,Oconnor2011,Dominik2012,Mapelli2016,ArcaSedda2018a,
Stevenson2019,Spera2019,Olejak2021b,Vanson2021,Ford2021,Broekgaarden2022}). Many of the above studies  
present BH-BH merger masses, but the evolution of mass with redshift is usually not shown nor discussed 
in detail. One notable exception is work by ~\cite{vanSon2022} who present detailed analysis of primary 
(more massive) BH mass and its evolution with redshift for BH-BH mergers for two (RLOF and CE) sub-channels 
of isolated binary evolution. None, however, have focused on average total BH-BH mass and its evolution 
with redshift. Here we present and discuss the cosmic evolution of average 
total mass of BH-BH mergers arising from various evolutionary channels. We note that 
empirical studies have been performed to assess changes in the BH-BH mass distribution shape with 
redshift, but thus far no definitive evidence of such mass distribution evolution has been found 
\citep{Fishbach:2021yvy,TiwariFeaturesInBBH,GWTC3b}.

\section{Models of BH-BH formation}
\label{sec:models}

\subsection{Isolated Binary Evolution Models}

We use three models of BH-BH formation from isolated binary stars that were calculated with the 
{\tt StarTrack} population synthesis code \citep{Belczynski2002,Belczynski2008a}. The standard 
evolution involves evolution of massive binary stars with empirically derived initial conditions 
\citep{Sana2012,deMink2015}. In particular, for massive stars the initial mass function (IMF) is 
adopted to be a power-law ($\propto M^{-2.3}$) and it extends to initial star mass of $150\msun$.
Input physics includes wind losses for massive stars: O/B star and WR-type winds in addition to LBV 
winds \citep{Vink2001,Belczynski2010b}. We treat the accretion onto compact objects during RLOF and 
from stellar winds using approximations presented by \cite{King2001,Mondal2020}. CE evolution is 
treated with the energy balance prescription of \cite{Webbink1984} and we limit accretion onto 
compact objects during a CE phase to $5\%$ of the Bondi rate \citep{MacLeod2017a}. 
We employ two models of core-collapse supernova (SN) engine in NS/BH mass calculation
~\citep{Fryer2012} that either allow (delayed SN engine) or do not allow (rapid SN engine) for 
filling in the lower mass gap (the dearth of compact objects with mass $\sim 2-5\msun$) with compact 
objects between NSs and BHs~\citep{Belczynski2012a,Zevin2020}. We assume the minimum BH mass of 
$2.5\msun$. In fact, we use the updated variants of rapid and delayed SN engines from 
~\cite{Olejak2022a}. This study allows for gradual filling of the lower mass gap with compact 
objects through a single parameter: $f_{\rm mix}$. Isolated binary models employ either 
$f_{\rm mix}=0.5$ (close to original delayed SN engine) and $f_{\rm mix}=4.0$ (close to the original 
rapid SN engine). For simplicity, when discussing binary models we will refer to them as having 
delayed or rapid SN engine physics.
Pulsational pair-instability SNe (PPSN) and pair-instability SNe (PSN) are assumed to be weak and 
allow for the formation of BHs with mass up to $\sim 55\msun$ \citep{Leung2019,Belczynski2020b}. We 
use the fallback-decreased NS/BH natal kicks with $\sigma=265\kms$ \citep{Hobbs2005}, and do not 
allow CE survival for Hertzsprung gap donors \citep{Belczynski2007}. 
The results of population synthesis calculations are post-processed with ~\cite{Madau2017} star 
formation rate history in Universe and with evolving with redshift metallicity distribution for 
Population I/II stars as presented in ~\cite{Belczynski2020b} to allow calculation of BH-BH merger 
population for redshift range $z=0-15$.

The standard physics presented above produces a majority of BH-BH mergers through the classical (CE) 
isolated binary evolution channel. In this channel, binaries that undergo RLOF with a donor star that 
is approximately twice the mass of the accretor or more ($q\gtrsim 2$) are subject to CE evolution 
that leads to orbital decay, and may lead to BH-BH merger under favorable conditions. We adopt 
model M380.B from ~\cite{Olejak2021a} that produces $99.5\%$ of BH-BH mergers through CE evolution. 
Delayed SN engine is employed in this model.

In our second model, we adopt non-standard CE development criteria that allows for stable RLOF to 
proceed even for systems with donors as massive as eight times mass of the accretor ($q\lesssim8$; 
\citealt{Pavlovskii2017}). Such evolution, also under favorable conditions, may lead to orbital decay 
and formation of BH-BH merger \citep{Heuvel2017}.  We adopt model M480.B from ~\cite{Olejak2021a} 
that produces $94\%$ of BH-BH mergers through stable RLOF evolution (and the rest through CE).
Delayed SN engine is employed in this model.

In our third model we adopt standard CE development criteria, but extend the IMF to $200\msun$ and 
allow for less restrictive PPSN/PSN limits on BH mass. It was argued that it is possible that the 
upper BH mass gap is located in a different place on BH mass spectrum or may not exist at all 
\citep{Farmer2020,Costa2021,Farrell2021}. We adopt a model presented by \cite{Belczynski2020c} that 
avoids PPSN mass loss for helium core masses up to $M_{\rm He}<90\msun$, with stars above this mass 
threshold destroyed by PSNe. This allows for formation of BHs with mass up to $90\msun$ from 
low-metallicity stars. Delayed SN engine is employed in this model.

In our fourth model we adopt exactly the same input physics as in the third model, but we employ the 
rapid SN engine instead.

This sequence of models with specific choices of input physics was design to show the 
dependence of total BH-BH merger mass on some of the most important and most uncertain parts of 
stellar and binary physics employed in our isolated binary evolution models.

\subsection{Globular Cluster Dynamical Models}

We use globular cluster models from the BH-BH formation channel analysis of \cite{Zevin2021b}, which 
have data relevant to this work publicly available on Zenodo~\citep{Zevin2021_data_release}. These 
models originate from a grid of $96$ $N$-body models of collisionless star clusters simulated using 
the H\'enon-style cluster Monte Carlo code \texttt{CMC}~\citep{Henon1971,Henon1971a,Joshi2000,
Pattabiraman2013}, as described in \cite{Rodriguez2019}. Cluster birth times and cluster initial 
metallicity follow cosmological models for globular cluster formation~\citep{ElBadry2019}, as outlined 
in \citep{Rodriguez2018} and \cite{Rodriguez2018a}. In particular, clusters with three metallicities 
are used $Z=0.005,\ 0.001,\ 0.0002$. The initial binary fraction for stars born in these cluster models is 
set to $10\%$ and is independent of stellar mass; the true value for the binary fraction of stars 
born in globular clusters is observationally unconstrained.

Stars and binaries that are born in GCs are evolved with rapid evolutionary formulas 
\citep{Hurley2000,Hurley2002} with updated prescriptions for stellar winds, compact object masses, 
and supernova natal kicks~\citep[see][]{Rodriguez2016b,Rodriguez2018b}. Initial masses of single 
stars and primaries  in binary systems are taken from a Kroupa IMF ($\propto M^{-1.3}$, in the range 
$0.08-0.5\msun$ and $\propto M^{-2.3}$, in the range $0.5-150\msun$), with secondary components
in binaries drawn from a uniform mass ratio distribution. Semimajor axes of initial binaries are 
drawn flat in log ($P(a)da \propto 1/a$) between the point of stellar contact and the hard-soft 
boundary, where the orbital velocity of the binary equals the typical velocity of particles in the 
cluster (for a typical velocity of $10\kms$ this translates to $\sim 1-5 \times 10^5\rsun$ for 
equal-mass binaries with component mass between $30 \msun$ and $150 \msun$). 
Eccentricities are drawn from a thermal distribution ($p(e) de = 2e de$). Stellar winds 
for massive stars are adopted from \cite{Vink2001} during the main sequence and from \cite{Vink2005} for 
naked helium stars, with later evolutionary stages following \cite{Belczynski2010b}. Compact object 
masses are calculated based on 1-D supernova models that allow for fallback and direct BH formation, 
with a rapid supernova engine that produces a lower mass gap between NSs and BHs \citep{Fryer2012}. 
Note that for isolated binary evolution BH-BH formation we employed models with delayed SN 
engine. However, as we show in Section~\ref{sec:evolution} the choice of SN engine model does not 
impact average total BH-BH merger mass and its evolution with redshift.
Massive stars (with He cores above $45\msun$ are subject to PPSN/PSN and stellar-origin natal BH 
mass is limited to $\lesssim 45\msun$. BH natal kicks are taken from a Maxwellian distribution with 1-D 
$\sigma=265\kms$, and kicks are decreased by fallback (i.e., no natal kicks for massive BHs that form 
through direct core collapse). This set of models assumes that BHs are born with near-zero spin due to 
highly efficient angular momentum transport (e.g., \citealt{Fuller2019b,Qin2018,Belczynski2020b}). 
Spin-up of massive stars (BH progenitors) by tidal interactions in close binaries is not taken into 
account. Stars ($90\%$) and primordial binaries ($10\%$) in each cluster form the first generation of 
BHs (with masses limited to $\lesssim 45\msun$). 

Dynamical interactions and also mergers of BH-BH systems formed from primordial binaries produce 
future generations of BHs. Obviously, these BHs can be more massive than the first generation of 
BHs. The second generation of BHs is typically formed with high spins ($a\sim 0.7$) due to 
the angular momentum of the pre-merger double compact object orbit. Spins of the BH remnant are 
calculated using the precession package~\citep{Gerosa2016}, as described in~\cite{Rodriguez2018a}.
Asymmetries in the BH components of the merger (unequal masses, spins) will lead to a dispersion of 
spin magnitudes around $a \sim 0.7$. Some second-generation BHs receive gravitational-wave recoil 
kicks that may remove BHs from clusters, though since the models we consider assume BHs are born 
with negligible spin, almost $\sim 60\%$ of second-generation BHs are retained since recoil kicks 
depend then entirely on mass ratio and are typically $\lesssim 100\kms$~\citep{Rodriguez2019a}.
Those that remain bound continue to contribute to cluster evolution and to GC BH-BH merger 
populations. Due to the relatively low escape velocities of globular clusters ($\sim 30-100\kms$), 
higher-generation merger products are typically ejected from the cluster via gravitational-wave 
recoil kicks due to their components having large spins as a result of their formation from previous 
BH-BH mergers; $\lesssim 0.1\%$ mergers containing a second-generation BH remain bound to the 
cluster~\citep{Rodriguez2019a}.

\subsection{Open Cluster Dynamical Models} 
\label{ocmodel}

In this work, the long-term N-body evolutionary model set of star clusters as described in
\citet{Banerjee_2021} is utilized. The computed models are described in detail in
\citet{Banerjee_2020,Banerjee_2020c,Banerjee_2021} and are summarized below. 

The model star clusters, initially,
have mass within $2\times10^4\msun\leq\mcl\leq10^5\msun$, size (half-mass radius)
$1{\rm~pc}\leq\rh\leq2{\rm~pc}$, and metallicity
$0.0001\leq Z \leq0.02$, and they are subjected to a solar-neighborhood-like
external galactic field. The initial models are composed of
ZAMS stars of masses $0.08\msun\leq m_\ast\leq150.0\msun$
that are distributed according to the standard IMF \citep{Kroupa2001}.
About half of the models
have a primordial-binary population (overall initial binary fraction
$\fbin\approx5$\% or 10\%) where all O-type stars (stars
with ZAMS mass $\geq16\msun$) are paired among themselves
(i.e, initial binary fraction among O-type stars is $\fobin=100$\%)
following an observationally-deduced distribution of massive-star binaries
\citep{Sana_2011,Sana_2013,Moe2016}. Such cluster
parameters and stellar membership are consistent with those
observed in `fully'-assembled, (near-)spherical, (near-)gas-free young massive clusters 
(YMCs) which evolve into medium-mass open clusters (OCs) \citep{PortegiesZwart_2010}.
They continue to form, evolve, and dissolve in the Milky Way and other galaxies
(as such, anywhere in the Universe) as a part of the galaxies' ongoing star formation.

These model clusters evolve due to two-body relaxation \citep{Spitzer_1987},
close (relativistic) dynamical encounters (\citealt{Heggie2003},
without applying any gravitational softening),
and stellar evolution. This is achieved using $\nbseven$,
a state-of-the-art post-Newtonian (PN) direct N-body integration code
\citep{Aarseth_2012,Nitadori_2012} that couples with the
semi-analytical stellar- and binary-evolutionary scheme
$\bse$ \citep{Hurley2000,Hurley2002}. As detailed in \citet{Banerjee_2020},
the integrated $\bse$ is 
updated in regards to prescriptions of stellar wind mass loss
and formation of NSs and BHs. NSs and BHs form according to
the `rapid' or `delayed' core-collapse SN models of \citet{Fryer2012},
and the PPSN and PSN
models of \citet{Belczynski2016c}. The majority of the computed models of 
\citet{Banerjee_2021} employ the rapid-SN prescription although a few models 
employ the delayed-SN prescription, for exploratory purposes. The dynamical 
evolution and merger outcomes of the clusters are unlikely to be significantly
affected by this difference, as discussed in \citet{Banerjee_2020,Banerjee_2020d}.

A newly formed NS or BH
receives a natal kick that is modulated based on SN matter fallback onto it,
as in \citet{Belczynski2008a}. The adopted fallback model is that of \citet{Fryer2012}.
For all core-collapse SN remnants, the `base' natal kick is assigned from a Maxwellian 
distribution with one-dimensional dispersion of $\sigma=265\kms$, i.e., the same velocity 
dispersion as that of the Galactic single NS population \citep{Hobbs2005}.
However, due to momentum conservation,
the material fallback slows down the remnants, causing BHs of $\gtrsim10\msun$
to be retained in the clusters right after their birth. The material fallback
also shapes the mass distribution of NSs and BHs.
NSs formed via electron-capture SN (ECS; \citealt{Podsiadlowski2004})
also receive small natal kicks of a few $\kms$ \citep{Gessner_2018} and are 
retained in the clusters at birth. The present remnant scheme forms first-generation 
of BHs in the range $\approx 3-45\msun$ (the upper limit set by PPSN) and exhibits a PSN 
mass gap between $\approx 45-120\msun$ (see Fig.~2 of \citealt{Banerjee_2020}).
See \citet{Banerjee_2020} for further details and the description of the 
implementations in $\bse$ and $\nbseven$.

In $\nbseven$, the PN treatment is handled by the $\archain$ algorithm
\citep{Mikkola_1999,Mikkola_2008}. The treatment allows for
general relativistic (GR) evolution of the innermost NS- and/or BH-containing binary
of an in-cluster (i.e., gravitationally bound to the cluster)
triple or higher order compact subsystem, in tandem with the Newtonian-dynamical
evolution of the subsystem (Kozai-Lidov oscillation or
chaotic three-body interaction), leading
to the binary's (in-cluster) GR in-spiral and potential merger. The PN treatment
applies also to the GR evolution of in-cluster NS/BH-containing binaries that
are not a part of a higher-order subsystem. As demonstrated in previous
studies \citep{Banerjee_2010,Banerjee_2017,Banerjee_2018b,Rastello_2019,Banerjee_2020c},
the moderate density ($10^2-10^3$ stars pc$^{-3}$) and velocity dispersion ($\lesssim10\kms$)
in the model clusters make them efficient in dynamically
assembling PN subsystems, particularly, those composed of BHs.
On the other hand, the clusters' low escape speed ($\sim10\kms$) 
causes the dynamically-ejected BH-BH systems to be typically wide \citep{PortegiesZwart2000},
with a GR inspiral time \citep{Peters1964} often exceeding the Hubble time.
These cause the majority of the GR mergers from these computed clusters to be 
in-cluster BH-BH mergers.

In the computed models, star-star and star-remnant mergers happen
due to evolution of massive primordial binaries and three-body interactions. 
In the present models, a star-star merger results in the loss of
$\approx10\%$ of the total merging mass as suggested by hydrodynamical calculations
\citep{Lombardi_Jr__2002,Gaburov_2008}.
In a BH-star merger (forming a BH-Thorne-Zytkow Object, BH-TZO),
$\approx90\%$ of the star's mass is taken to be accreted
onto the BH.
The model grid used in this work comprises 74 long-term ($\sim10$ Gyr) evolutionary
cluster models (see Table~A1 of \citealt{Banerjee_2021}).

\section{Results}
\label{sec:results}

\begin{figure}
\hspace*{-0.4cm}
\includegraphics[width=0.5\textwidth]{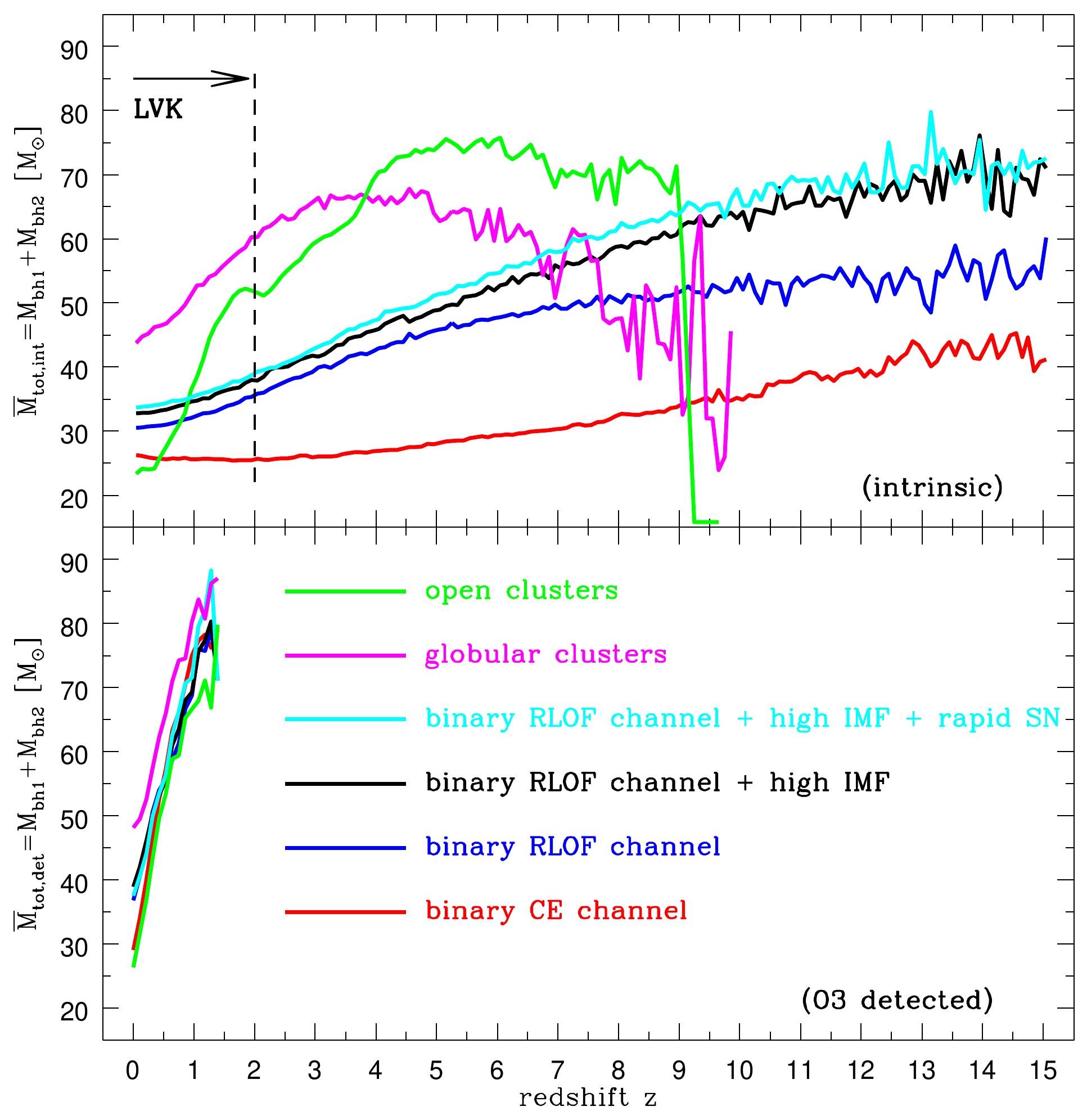}
\caption{
Top: intrinsic average total BH-BH merger mass change as a function of redshift for different BH-BH 
formation channels: globular cluster and open cluster dynamical channels, and CE and RLOF isolated 
binary evolution channels. For the RLOF channel we show three variations: first allows for stars to 
form to $150\msun$ and PPSN/PSN limit BH masses to $\sim 55\msun$, second allows for star formation 
to $200\msun$ and PPSN/PSN limit BH masses to $90\msun$, while third is the same as the second 
variation but employs rapid SN engine (all other isolated binary models employ delayed SN engine). 
We show the approximate redshift range for design (future) LIGO/Virgo/KAGRA detection of BH-BH 
mergers (the range for Einstein Telescope and Cosmic Explorer covers entire plot). 
Bottom: detected average total BH-BH merger mass as a function of redshift. The gravitational-wave
instrumental and observational selection effects are applied to the models from the top panel to show 
which BH-BH mergers may be currently (O3 sensitivity of LVK) detectable. Note that rather different 
intrinsic curves (top), become almost indistinguishable when detection biases are applied (bottom). 
}
\label{fig:bhmass}
\end{figure}

\subsection{Evolutionary Analysis}
\label{sec:evolution}

Our main result is shown in the top panel of Figure~\ref{fig:bhmass}. It shows evolution of the 
average intrinsic (not redshifted nor detection weighted) total BH-BH mass 
($\overline{M}_{\rm tot,int}=M_{\rm bh1}+M_{\rm bh2}$) with redshift ($z$). The evolution is shown for 
the six models (see Sec.~\ref{sec:models}) of BH-BH merger formation: the globular cluster dynamical 
channel, the open cluster dynamical channel, the classical CE isolated binary evolution channel, the 
RLOF isolated binary evolution channel, the extended RLOF channel (with more massive stars and BHs), 
and the extended RLOF channel with rapid SN engine. 

Within the approximate LIGO-Virgo-KAGRA design redshift range for detection of BH-BH mergers 
($z<2$)\footnote{The maximum detectable redshift of a 50$M_{\odot}$-50$M_{\odot}$ BH merger with 
Advanced LIGO at design sensitivity is $z_h=1.9$, as calculated with the gravitational wave distance 
calculator described in \cite{ChenHolzDistanceGW}.}, the CE channel produces the lowest mass BH-BH 
mergers ($\overline{M}_{\rm tot,int}\sim 25\msun$, irrespective of the redshift), the three RLOF 
channels produce more massive BH-BH mergers that show a slow increase of the average mass with 
increasing redshift as the metal-poor environments at higher redshifts produces more massive BHs 
($\overline{M}_{\rm tot,int}\sim 30\rightarrow 35\msun$ for $z=0\rightarrow 2$), and the 
globular cluster channel produces the most massive BH-BH mergers with fast increase of the average 
mass with redshift ($\overline{M}_{\rm tot,int}\sim 45\rightarrow 60\msun$ for $z=0\rightarrow 2$).
The open cluster channel generates the most dramatic evolution of the intrinsic average total BH-BH 
merger mass ($\overline{M}_{\rm tot,int}\sim 25\rightarrow 50\msun$ for $z=0\rightarrow 2$).

In the entire presented redshift range ($z<15$) that will be accessible to Einstein Telescope and 
Cosmic Explorer, the four isolated binary evolution channels produce gradually more massive BH-BH 
mergers with increasing redshift (at $z=15$ we find: $\overline{M}_{\rm tot,int}\sim 40\msun$ for 
the CE channel, $\overline{M}_{\rm tot,int}\sim 55\msun$ for the RLOF channel, and 
$\overline{M}_{\rm tot,int}\sim 70\msun$ for both RLOF channels with extended IMF). For BH-BH 
mergers in the globular cluster channel, we find that the average mass first increases with 
redshift (reaching a maximum of $\overline{M}_{\rm tot,int}\sim 65\msun$ at $z=3-5$) and then 
gradually decreases with increasing redshift. For the open cluster channel, we find that the average 
BH-BH mass first increases with redshift, reaching high values 
($\overline{M}_{\rm tot,int}\sim 70\msun$ at $z=4-9$) and then sharply decreases for higher 
redshifts. 

The general increase of the intrinsic average BH-BH mass with redshift for all four isolated binary 
evolution channels is easily understood in the framework of cosmic chemical evolution. The chemical 
composition of stars (i.e., metallicity $Z$) is the primary factor setting the mass of stellar-origin 
BHs \citep{Belczynski2010b}. At lower metallicities, stellar winds are weaker (which removing mass 
from stars), and more massive BHs are able to form. Since metallicity decreases with redshift, more 
massive stellar-origin BHs form at higher redshifts. There is an exception from this general trend 
for the CE BH-BH formation channel at low redshifts ($z\lesssim 2-3$), where the BH-BH merger total mass 
remains constant and does not change with redshift. We explain this behavior in Section~\ref{sec:appA}.

The lowest average total BH-BH merger mass (at any redshift) is found for the CE BH-BH binary formation 
channel. The actual value of the average mass is a combination of the adopted stellar wind mass loss for 
massive stars and their dependence on metallicity, the cosmic evolution of metallicity with redshift, 
and the specific evolutionary sequences (set by stellar/binary input physics assumptions) producing 
BH-BH mergers in the CE binary channel. The RLOF binary channel produces on average more massive BH-BH 
mergers. Since for this channel wind mass loss and cosmic metallicity evolution are the same as for 
the CE channel, the entire difference in average total mass originates from different stellar/binary 
evolutionary assumptions. We explain the (non-trivial) origin of this difference in Section~\ref{sec:appB} 
Finally, the two RLOF models with high (extended) IMF naturally produce the most massive BH-BH mergers 
among our isolated binary models. In these two models, BHs may form up to $90\msun$ as compared to maximum 
mass of a BH of $\sim 55\msun$ in the other two binary models. We note slightly larger total BH-BH 
masses ($\lesssim$  few $\msun$) for model with the rapid SN engine as the BHs that would form in the 
lower mass gap in the delayed SN engine model are simply forming with masses just above the mass gap. 
Note that the small differences between these two models with extended IMF but with different SN engines 
are much smaller than differences associated with the BH-BH formation mode (CE versus RLOF channels) or 
the change of IMF and PPSN/PSN model. 

The globular cluster channel produces on average significantly more massive BH-BH mergers at low and 
intermediate redshifts ($z<5$) compared to any presented isolated binary evolution model by about 
$\overline{M}_{\rm tot,int}\sim 10-20\msun$ (see Fig.~\ref{fig:bhmass}). This is due to the most massive BHs 
in the cluster more readily synthesizing binaries through dynamical interactions, as well as the 
contribution of mergers containing a second-generation BH that was retained in the cluster following 
its first merger. The average total merger mass drops at high redshifts ($z\gtrsim 7$) when GC BH-BH 
mergers become more dominated by BH-BH mergers from primordial binaries. 

Unlike the isolated evolution channels considered, the globular cluster model has a peak in the average 
mass of BH-BH mergers near the peak of globular cluster formation, at redshifts of $z\sim 3-5$. The most 
massive BHs in the cluster ($\gtrsim 20\msun$) are typically the first to be dynamically processed, 
either merging in the cluster at early times or being ejected and inspiraling over longer timescales. 
Less massive BHs born in the clusters ($\sim 10-15\msun$) follow suit and typically merge at 
later times than their more massive counterparts, leading to the decrease in the average total mass lower 
redshifts ($z\lesssim 3$). At $z\gtrsim 5$, primordial binaries in the cluster that evolve and lead to 
BH-BH mergers formed in quasi-isolation (without much assistance from dynamical interactions) have a 
larger contribution to the total population of BH-BH mergers, driving the overall decline in the 
average mass at large redshifts. These systems rapidly merge on timescales shorter than the mass 
segregation timescale in their GC hosts of $\mathcal{O}(100\,\mathrm{Myr})$. However, dynamically 
processed mergers still occur at these higher redshifts, which combined with small number statistics 
(GCs are just beginning to form) leads to the fluctuating behavior in the average total mass evolution at 
$z\gtrsim 5$. The relative contribution of BH mergers from quasi-isolated evolution prior to the 
peak of globular cluster formation and their impact on the average mass change with redshift depends 
on the metallicity evolution of cluster formation, which is distinct from the metallicity evolution 
of isolated binaries in galactic fields.  

Interpretation along lines similar to the case of GCs applies to the evolution of the average total mass
with merger redshift for OCs (Figure~\ref{fig:bhmass}; top panel). However, a twist
for OCs is that such systems, being 1-2 orders of magnitude less populous than but of similar (initial) length
scale as GCs, have a factor of a few to tens shorter two-body relaxation time \citep{Spitzer_1987} compared
to GCs. In OCs, the most massive BHs/BH-progenitors segregate to the cluster center in $\lesssim10$ Myr
\citep[see][]{Banerjee_2022}.
Also, before the formation of BHs, a fraction of the BH progenitors already undergo complete
mass segregation due to the shorter relaxation/mass-segregation timescale.
These cause the dynamical processing and mergers among BHs to commence at earlier
ages (i.e. at larger redshifts) and closer to the quasi-isolated mergers in
OCs than in GCs. Also, due to the smaller member number and hence primordial-binary
population in OCs, the models produce a much smaller number of quasi-isolated
BH-BH mergers per cluster than the GC models \citep{Fragione_2020b}.

The effects of these differences in OCs can be seen in Figure~\ref{fig:bhmass}.
As opposed to GCs, $\overline{M}_{\rm tot,int}$ for OCs rises sharply at
$z\approx9$ (fewer quasi-isolated mergers), thereafter remaining nearly flat and
peaking weakly in between $5 < z < 6$ (pre/faster mass segregation), and drops off
towards low $z$ faster than GCs (faster dynamical processing due to shorter
relaxation times in OCs). It is also to be noted that the OCs are taken
to follow a different metallicity-dependent cosmic star formation history 
(\cite{Chruslinska2019b} their 'moderate-Z' relation; see \cite{Banerjee_2021} 
for the details) than those for both the 
isolated-binary and GC models presented here, which would also cause differences 
in the redshift dependence of $\overline{M}_{\rm tot,int}$. Although OCs are inefficient 
in producing massive BHs via hierarchical mergers, the computed OC models adopt a 
small, $\approx10$\%, mass loss in star-star mergers (based on hydrodynamic stellar 
merger calculations; see Sec.~\ref{ocmodel}), which is the main channel for forming 
the most massive BHs lying within the PPSN/PSN mass gap in these models 
\citep{Banerjee_2020c,Banerjee_2022}. Furthermore, a $90\%$ BH-TZO accretion is 
adopted (Sec.~\ref{ocmodel}). These, along with the tendency of having lower 
metallicity at high redshifts than globular cluster or binary models, results in a 
higher maximum $\overline{M}_{\rm tot,int}$ for OCs (Figure~\ref{fig:bhmass}).

\subsection{Comparison to Observed Gravitational Waves}
\label{sec:data}

Although all models explored in this work show notable differences in the average total mass of 
mergers and its evolution with redshift (the only exception being the two RLOF models with extended 
IMF: one with rapid and one with delayed SN engine), 
there is no guarantee these differences can be measured with current gravitational-wave sensitivity. 
Firstly, gravitational-wave sources are subject to 
strong selection effects, which not only limit the distance to which BH-BH mergers
can be detected, but also preferentially enable detection of certain binary mass and spin 
configurations. As a result, the properties of the detected population of BH-BH mergers
will not match those of the intrinsic population. Furthermore, gravitational-wave 
detectors are subject to a variety of sources of noise, so the true masses, spins, 
and distances of detected events can only be estimated, often with sizable uncertainties. 
Finally, we are still in the era of small number statistics when it comes to gravitational-wave detections,
so even if the true properties of a detected system were known, there still exists a counting
uncertainty in the number of similar events occurring in the Universe.

\begin{figure}
\hspace*{-0.4cm}
\includegraphics[width=0.5\textwidth]{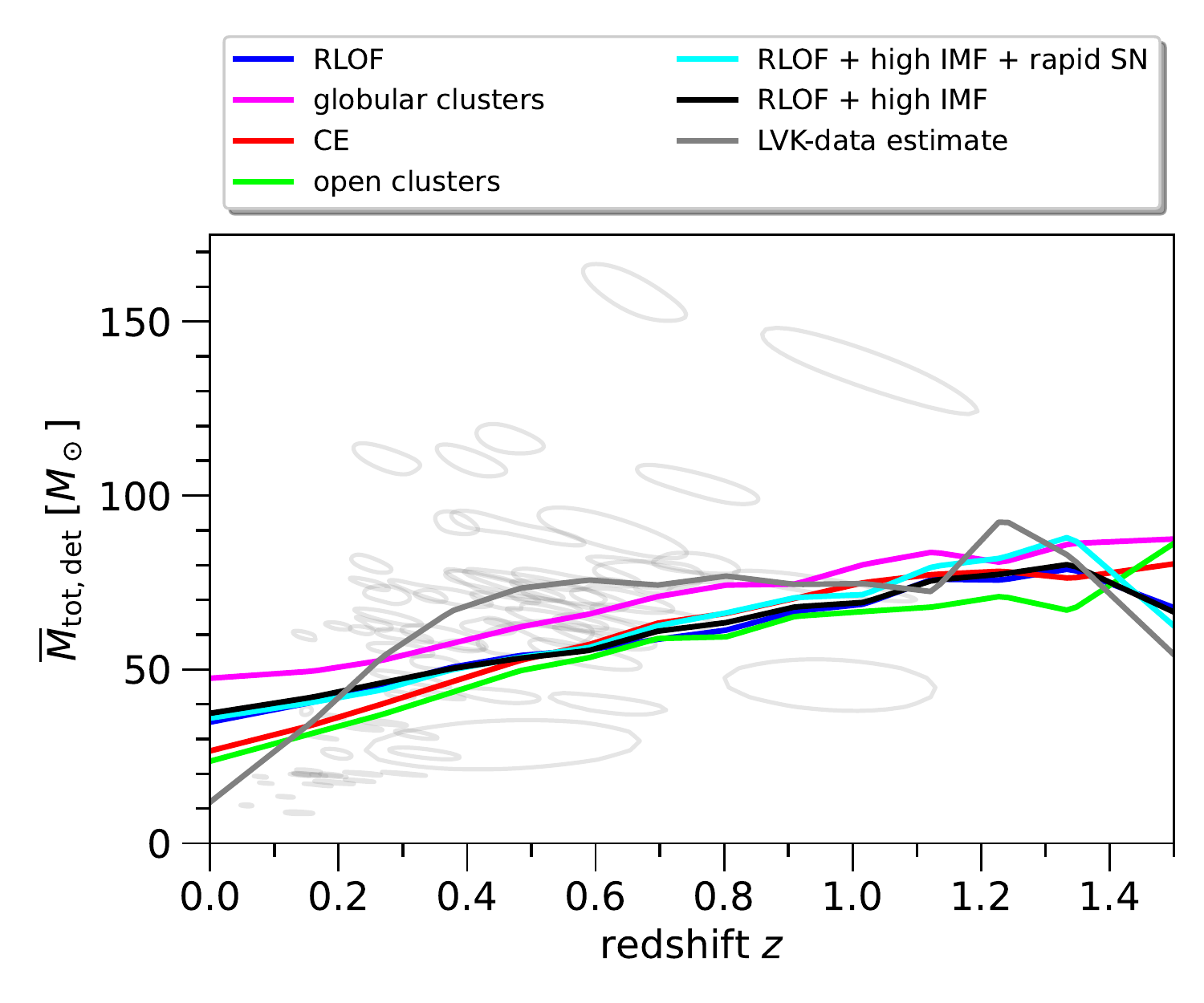}
\caption{The average detected total mass as a function of redshift for the six models considered 
in this work. The models are overlaid on the gray $90\%$ credible interval parameter regions for 
BH-BH mergers from LVK O3, re-weighted to an astrophysically realistic prior. The gray line shows 
an estimate of the mean detected total mass of these events in different redshift bins for all
posterior samples under the aforementioned prior.
}
\label{fig:avgdetmass}
\end{figure}

Given these combined effects as well as the poorly understood systematic uncertainties in the six 
population synthesis models presented here, we take a simple approach to assessing whether these 
models are distinguishable with current detections, and whether any of them are a reasonable fit to 
the data, in analog to the investigations in \citet{Fishbach:2021yvy}. In Figure~\ref{fig:avgdetmass}, 
we show the average {\it detected} mass $\overline{M}_{\rm tot, det}$ as a function of redshift under 
each of our six models by computing the average detected total mass in 14 equally spaced redshift 
bins (i.e.~15 bin edges) from $z=0$ to $z=1.5$. These curves are also reproduced in Figure~\ref{fig:bhmass} 
for comparison to the intrinsic average masses. $\overline{M}_{\rm tot, det}$ is essentially 
$\overline{M}_{\rm tot,int}$ filtered through the LVK detector and analysis selection function.  
To simulate these effects of the detector and analysis selection bias on the detected
populations, we re-weight the detected software injections provided in \citet{GWTC3b} 
by the intrinsic simulated populations. To perform the re-weighting, a histogram of the 
population synthesis samples from each model is used to approximate the 
intrinsic simulated population probability distribution over masses and redshifts. 

The model curves in Figure~\ref{fig:avgdetmass} show the average detected total mass increasing as a 
function of redshift, which is expected given that less-massive sources are detectable
out to lower redshifts than the more massive sources. 
The rate of change of the average detected total mass with redshift is approximately
the same for all our models, though the CE channel shows the strongest increase with redshift.
To compare these curves to those in Figure~\ref{fig:bhmass}, consider that
the slopes of the $\overline{M}_{\rm tot,det}$ lines are strongly influenced by the selection effects, which will 
increase the lowest detectable mass with increasing redshift and 
therefore force models with low, constant intrinsic average
mass like the CE channel to appear to evolve with redshift in the detected population.
In contrast, for the open clusters, the average intrinsic mass changes from $\sim 25$ to $\sim 50 M_\odot$ 
between redshift 0 and 1 which matches what is seen in the detected average mass.
This means that the true average mass evolution with redshift 
in the open cluster model is roughly following the O3 LVK selection effects. 
The similarity between slopes of the $\overline{M}_{\rm tot, det}$ curves 
suggests that the changes in the total mass distribution mean over redshift is 
not a good distinguisher of these models given current sensitivity. 
However, the overall mass scale is more telling: the GC model predicts a higher
average detected mass than the other models by $\sim10-20M_\odot$, regardless of redshift. 

To compare these simulated average detected masses to the masses of detected BH-BHs, in gray we overlay 
the $90\%$ credible regions on detector-frame total mass and redshift for the BH-BH systems in GWTC-3
from the third LVK observing run \citep{Abbott2021a,Abbott2021b,GWTC3a}. Only BH-BH mergers from O3 are 
selected for uniform sample. The credible regions are generated via kernel density estimates (KDEs) of 
posterior samples provided by the LVK, re-weighted to a prior that approximates empirical 
measurements of the merger population distribution \citep{GWTC3b,Roulet2021,LIGO2020a}. Specifically, 
we use (1) a broken power law prior on the more massive BH in the binary's mass, with 
$p(m_1)\propto m_1^{-1.7}$ for $m_1<40M_\odot$, and $p(m_1)\propto m_1^{-5}$ for $m_1\geq40M_\odot$,
(2) a $p(q)\propto q$ prior for the mass ratio $q$, and 
(3) a uniform-in-comoving-time-volume prior on $z$, i.e.~$p(z)\propto \frac{dV}{dz}\frac{1}{1+z}$. 
Additionally, we estimate the mean mass of these detections over redshift by
computing a single total-mass mean from the combined samples of all events in the same redshift 
bins used for the six models' curves. We show this estimate in gray in Figure~\ref{fig:avgdetmass}. 

At $z\sim0.5$, all the models appreciably under-predict the estimated mean detected mass by 
$\sim 25 M_\odot$, except GC, which underpredicts by $\sim 10 M_\odot$. This is consistent with the 
fact that the GC model is the only model with a non-negligible fraction ($>1\%$) of systems with 
$M_{\rm tot,int}>100M_\odot$ below $z=1$. At the highest redshifts, it is not clear which model(s) may 
fit the data well given that there are few events detected at these distances, and those events have 
large uncertainties and low significance. At redshifts less than $\sim 0.2$, all the models 
over-predict the average detected mass. This is simply because none of the models predict a 
sufficiently high rate of low-mass mergers. Overall, these findings, especially at low masses and 
redshifts, suggest that none of these models alone (or even a mixture of them), are a good fit to 
GWTC-3 events. However, the GC model offers the closest $\overline{M}_{\rm tot, det}$ prediction
for high redshift events, but would require either a different set of input physics or another 
channel also with a different set of input physics than explored in this work to fill in the low and 
high ends of the mass distribution. 

We emphasize that the explorations here are primarily illustrative. None of the models
herein have been tuned to fit the LVK data, nor have we performed a simultaneous
fit to all binary parameters in the LVK catalog accounting for measurement
uncertainties. However, comparisons such as the one shown in Figure~\ref{fig:avgdetmass}
will be invaluable for assessing the viability of BH-BH formation
channel models and enable a narrowing of the parameter space
for easier follow-up studies that perform a complete statistical analysis. 

\section{Conclusions}
\label{sec:conclusions}

We have employed several BH-BH merger formation models to show evolution of the average intrinsic 
and detected total BH-BH merger mass with redshift. Our main result is shown in Figures~\ref{fig:bhmass} 
and \ref{fig:avgdetmass}. In our analysis we have included the classical isolated binary 
CE evolution formation channel, three stable RLOF channels: one with a standard IMF (stars with masses 
$<150\msun$ and PPSN/PSN limiting BH masses to $<55\msun$), and two with an extended IMF (stars with 
masses $<150\msun$ and PPSN/PSN limiting BH masses to $<90\msun$) but with different SN engines 
(delayed and rapid), and globular cluster and open cluster dynamical BH-BH formation channels. 

The model differences in average intrinsic BH-BH total mass are both quantitative and qualitative. The 
lowest average masses are produced by the CE binary channel ($\overline{M}_{\rm tot,int}\sim 25\msun$ at 
$z=0-2$), with somewhat larger masses found for RLOF channels ($\overline{M}_{\rm tot,int}\sim 30\msun$ 
at $z=0$ and increasing to $\sim 35\msun$ at $z=2$). The largest masses in LIGO/Virgo/KAGRA range are 
produced by globular clusters ($\overline{M}_{\rm tot,int}\sim 45\msun$ at $z=0$ with notable increase 
to $\sim 60\msun$ at $z=2$). Open cluster BH-BH mergers show the most pronounced evolution of the average 
total mass in this redshift range ($\overline{M}_{\rm tot,int}\sim 25\msun$ at $z=0$ and $\sim 50\msun$ 
at $z=2$), bridging isolated binary and globular cluster models. 

These differences become more pronounced for higher redshifts. Isolated binary channels show monotonic 
increase of the average intrinsic BH-BH merger mass with redshift, with the exception that for the CE 
channel the average intrinsic mass is constant for low redshifts ($z\lesssim 2-3$) and then increases with 
increasing redshift. A different behavior is found for the globular cluster model, for which the
average intrinsic merger mass increases with redshift to $z\sim 4$, and then decreases with redshift. 
The open cluster model shows similar behavior to the globular cluster model, however the largest intrinsic total 
BH-BH masses are produced at higher redshifts ($\overline{M}_{\rm tot,int}\gtrsim 70\msun$ for $z\sim 5-9$). 
These trends are understood in the framework of input physics employed in the models (see Sec.
~\ref{sec:evolution} for details). 

We have shown that not only the formation channel (e.g., isolated binary versus cluster BH-BH formation)
significantly affects intrinsic average BH-BH merger mass, but that some parts of highly uncertain input 
physics within a given formation channel have big impact. For example, changing CE development criteria 
for isolated binary evolution channel significantly impacts BH-BH merger formation mode (CE versus RLOF) 
and the intrinsic total BH-BH merger mass. Similar statement can be made about impact of treatment of 
PPSN/PSN and the extend of IMF in binary evolution calculations. However, not all input physics is that 
important for the intrinsic average total BH-BH merger mass and its evolution with redshift. For example,   
SN engine model that regulates the existence and numbers of compact objects with mass in the lower mass 
gap ($\sim 2-5\msun$) does not significantly impacts intrinsic average total BH-BH merger mass. Since all 
these parts of physics of BH-BH merger formation (CE, SN engine, PPSN/PSN) are highly uncertain we 
adopted models with various assumptions on parameters regulating these processes. 

The evolution of the averaged detected BH-BH merger total mass does not differ as much between formation 
channels compared to the intrinsic mass (see Fig.~\ref{fig:avgdetmass}). The four binary models 
and the open cluster model show a similar behavior for redshifts in which mergers are currently detected 
($z\lesssim 1$). Only the globular cluster model results in moderately heavier detected BH-BH mergers (by 
$\lesssim 10\msun$ for $z\lesssim 1$). This is caused by LVK detection biases that affect different models 
in different ways (as explained in Sec.~\ref{sec:data}). These differences in mass evolution
with redshift are too small to differentiate 
between models with the current LVK data due to large detection errors. 
Additionally, none of the employed models matches the average detected BH-BH merger masses over entire BH-BH 
detectable redshift range. However, the adopted globular cluster model matches the current data better
than our isolated binary or open cluster models at moderate redshifts $z\sim0.5$.

The physics of BH-BH merger progenitors for isolated binary, globular cluster, and open cluster formation 
channels is still subject to many uncertainties \citep{Belczynski2022,Santoliquido2021,Chatterjee2017,Conroy2010}. 
Major overall uncertainties include the underlying cosmic star formation rate of isolated binaries and 
stars/binaries in globular and open clusters, and metallicity evolution with redshift \citep{Sharda:2021} 
within each formation channel \citep{Neijssel:2019}. Basic stellar evolution uncertainties include fusion 
reaction rates \citep{Fields2018}, mixing in stellar interiors \citep{Zhang2013}, stellar wind mass loss 
rates \citep{Bjorklund2022}, radial expansion, and core-collapse/supernova physics \citep{Schneider2019} 
associated with compact object formation \citep{Belczynski2022}, as well as initial conditions such as the 
initial mass function, mass ratio, orbital separation, and eccentricity \citep{Klencki:2018zrz}. Binary 
evolution physics is highly uncertain in its treatment of stellar interactions, whether these are stable or 
unstable (CE) RLOF events \citep{Ablimit:2018, Dominik:2012}. In modeling of dynamical BH-BH formation 
unknown initial conditions of globular and open clusters (size, density, binary fractions) along with 
uncertain outcomes of stellar mergers and BH (natal and gravitational-wave) kicks hinders their predictions. 

Keeping these uncertainties in mind, we can deliver only a very limited conclusion. For some choices 
of input physics in a given BH-BH formation channel (six presented models) we predict that the 
average intrinsic total BH-BH merger mass evolution with redshift changes from model to model, 
sometimes in rather significant ways. However, due to BH-BH detection selection effects, small 
statistics of detected sources, and significant uncertainties in detected BH-BH masses and redshifts, 
the current LVK O3 data does not enable us to identify any model as fitting the data better than another. 
Additionally, no models (or their potential combinations) fit the entirety of the data well due to their 
dearth of low-mass events relative to the detected population. More data from LVK in O4 and O5 may change 
this conclusion. At this moment, it is clear that the average total mass (and its evolution with redshift) 
may be very different for various formation models, and this may help in the future to eliminate some 
models and support others, and guide theoretical work on the physics of BH-BH merger progenitors. 
The total BH-BH merger mass and its evolution with redshift may provide, along other observables, an 
extra leverage in search for the BH-BH formation site/channel.

\vspace*{-0.2cm} \acknowledgements
KB and AO acknowledge support from the Polish National Science Center grant Maestro 
(2018/30/A/ST9/00050). ZD is supported by the CIERA Board of Visitors
Research Professorship. Support for MZ is provided by NASA through the NASA Hubble Fellowship grant 
HST-HF2-51474.001-A awarded by the Space Telescope Science Institute, which is operated by the 
Association of Universities for Research in Astronomy, Incorporated, under NASA contract NAS5-26555. DC is supported by the STFC grant ST/V005618/1.
SB acknowledges support from the Deutsche Forschungsgemeinschaft (DFG; German Research Foundation) through
the individual research grant ``The dynamics of stellar-mass black holes in dense stellar systems and their
role in gravitational-wave generation'' (BA 4281/6-1; PI: S. Banerjee). SB acknowledges the generous support
and efficient system maintenance of the computing teams at the AIfA and HISKP.
This work was initiated and performed in part at Aspen Center for Physics, which is supported 
by National Science Foundation grant PHY-1607611.
This material is based upon work supported by NSF's LIGO Laboratory which is a major facility fully funded by the National Science Foundation.

\bibliography{biblio}

\section{APPENDIX A}
\label{sec:appA}

All four binary models show a general increase in the average intrinsic total BH-BH merger mass with 
redshift (see Fig.~\ref{fig:bhmass}). However, there is an exception from this trend for the CE BH-BH 
formation channel at low redshifts ($z\lesssim 2-3$). Here, we provide an explanation for this behavior. 

\begin{figure}
\hspace*{-0.0cm}
\includegraphics[width=0.5\textwidth]{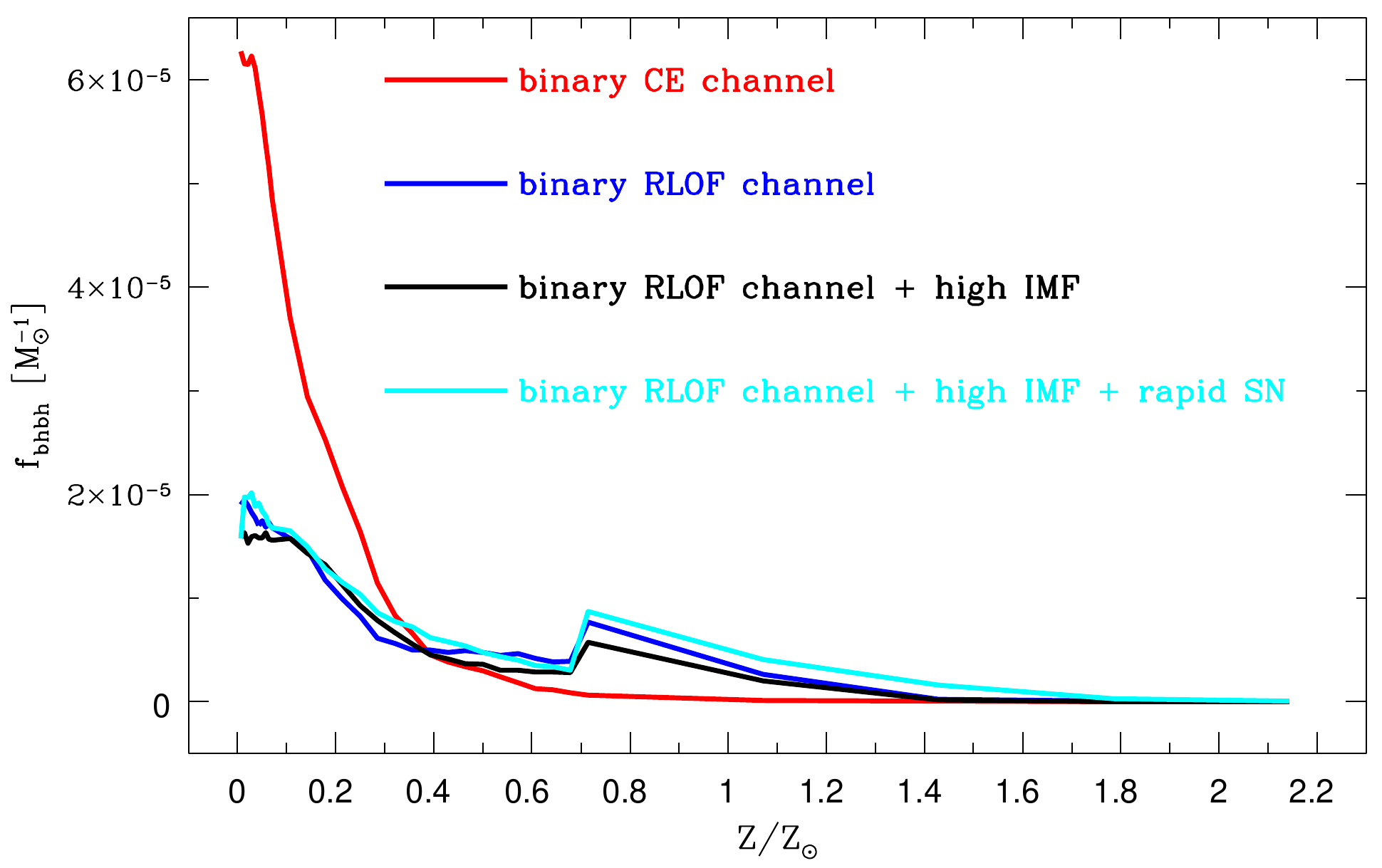}
\caption{
BH-BH merger formation efficiency per unit mass as a function of metallicity for the CE and
three RLOF isolated binary evolution channels.
}
\label{fig:efficiency}
\end{figure}

For this channel, BH-BH merger formation depends sensitively on radial expansion of BH progenitor 
stars and thus on metallicity \citep{Belczynski2010a}. For our CE model, we only allow for CEs to 
develop and successfully form close binaries for late donor evolutionary stages (e.g., during core 
He-burning). Significant radial expansion during core He-burning is found at low metallicity, as 
high metallicity H-rich stellar envelopes are usually severely depleted by stellar winds for massive 
stars. As metallicity of stellar populations increases with decreasing redshift, the CE channel BH-BH 
merger formation efficiency significantly drops as seen in Figure~\ref{fig:efficiency}. For low 
redshifts ($z\lesssim 2-3$) the average metallicity of stellar populations becomes significant 
($Z>0.5\zsun$, see Fig.7 of \cite{Belczynski2020b}) and the formation efficiency of BH-BH mergers is
very small. The low mass BH-BH mergers that form from high metallicity stars do not contribute 
significantly to overall population of BH-BH mergers at these redshifts. The majority of mergers at 
these redshifts originate from low-metallicity stars that still form in the tail of metallicity 
distribution even to $z=0$. Massive mergers that form at low metallicity at high redshifts do not 
contribute significantly at low redshifts as merger rates fall off steeply with delay time 
\citep{Dominik2012}. This sharp drop in BH-BH merger formation efficiency with metallicity 
(Fig.~\ref{fig:efficiency}) is responsible for the flat behavior of the average BH-BH merger mass at 
low redshifts. This does not affect RLOF channels as they depend on expansion of stars at much 
earlier evolutionary stages (e.g., during the Hertzsprung gap) which are not affected as much by stellar 
winds and metallicity.

\begin{figure}
\hspace*{-0.0cm}
\includegraphics[width=0.5\textwidth]{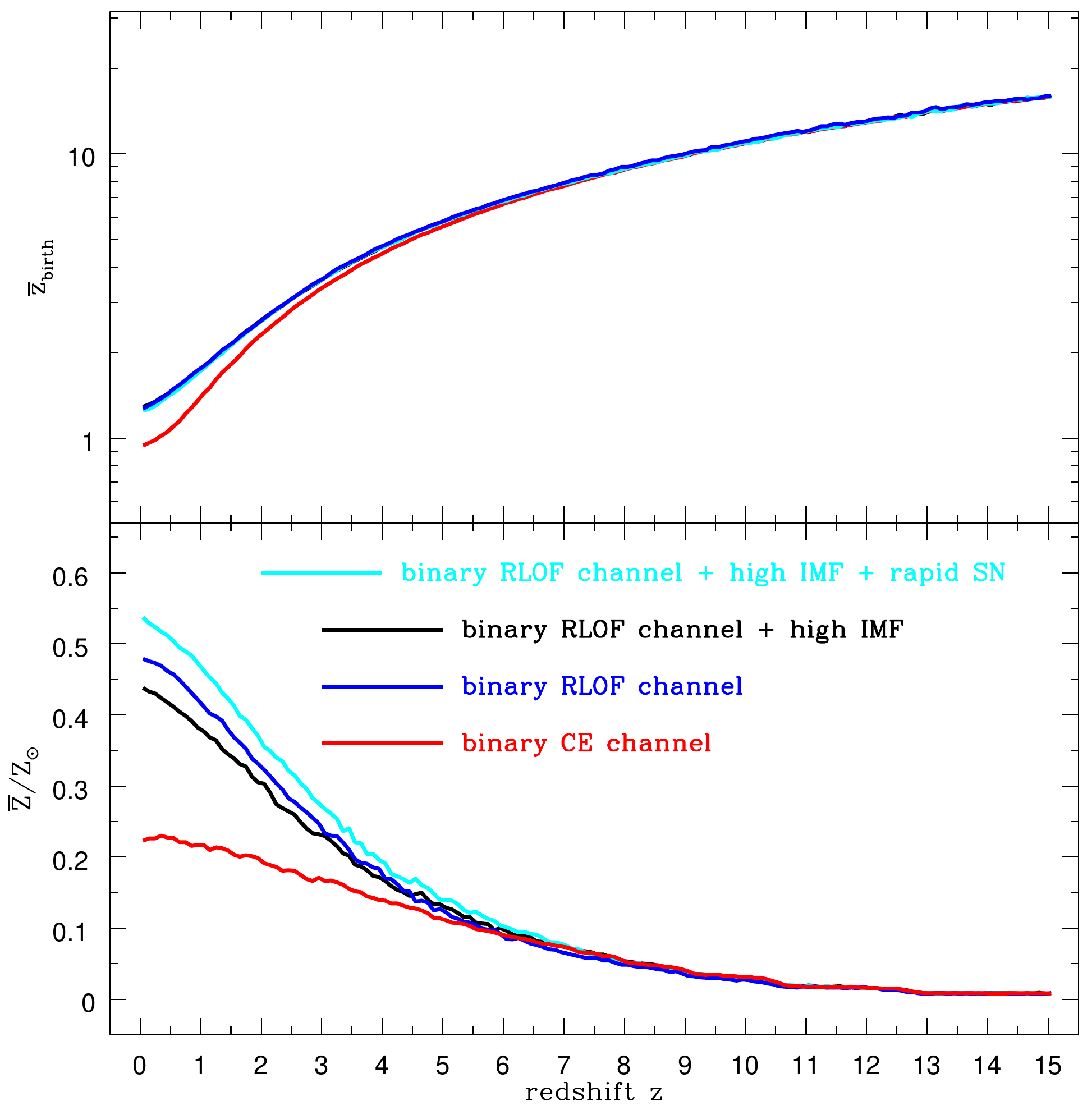}
\caption{
Top: average birth redshift of binary progenitors of BH-BH mergers that merge at a given 
redshift for four isolated binary formation channels. Note that black and cyan lines are right under 
blue line. Bottom: average metallicity of binaries that produce BH-BH mergers at a given redshift.
}
\label{fig:Zz}
\end{figure}

\section{APPENDIX B}
\label{sec:appB}

Here, we explain the difference in average total mass of BH-BH mergers between the CE formation 
channel and RLOF channels. It is clearly visible that RLOF channels produce much more massive 
BH-BH mergers at all redshifts (see Fig.~\ref{fig:bhmass}). 

\begin{figure}
\hspace*{-0.0cm}
\includegraphics[width=0.5\textwidth]{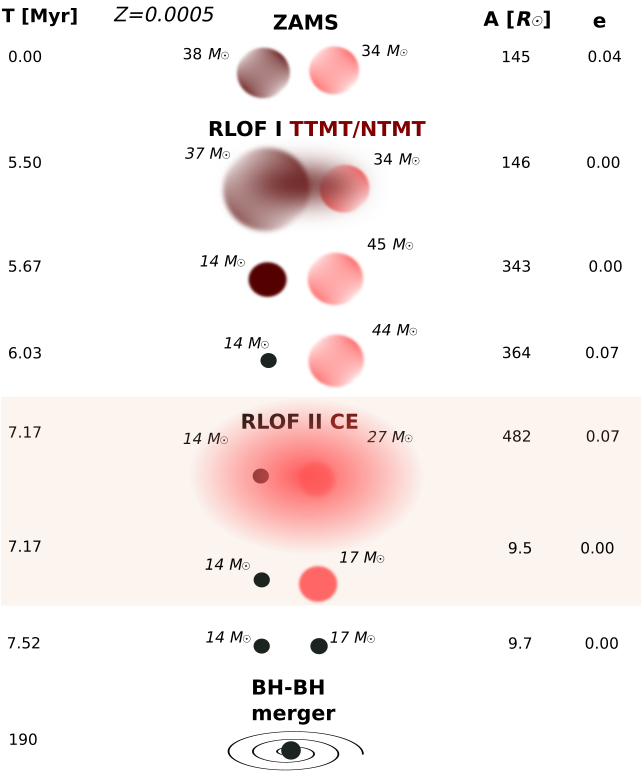}
\caption{
The formation scenario for a typical BH-BH merger at redshift $z \approx 6$ in the classical (CE) 
isolated binary evolution channel. The shaded part of the evolution shows the most important part of 
evolution that leads to orbital decay, allowing for the formation of close BH-BH binary that can 
merge in the Hubble time. 
}
\label{fig:evol1}
\end{figure}

\begin{figure}
\hspace*{-0.0cm}
\includegraphics[width=0.5\textwidth]{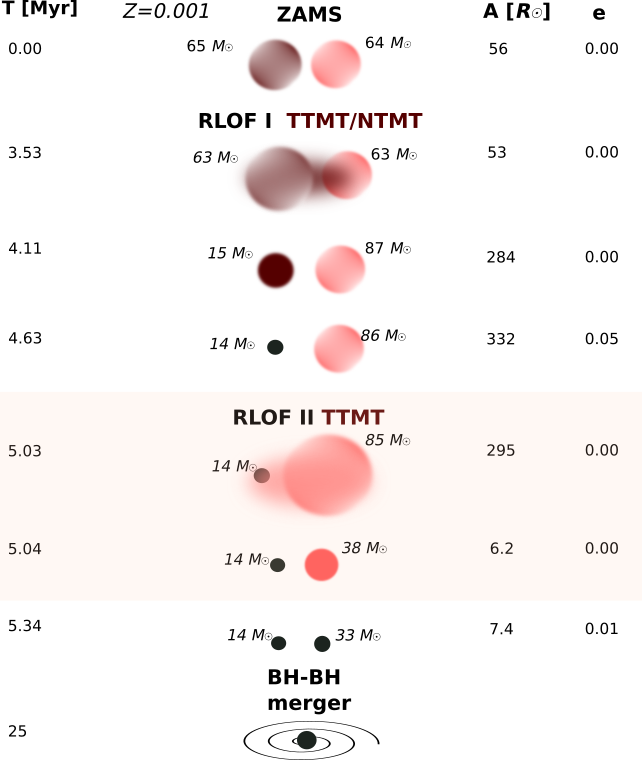}
\caption{
The formation scenario for a typical BH-BH merger at redshift $z \approx 6$ in the stable RLOF 
channel (standard IMF, delayed SN engine). The shaded part of the evolution shows the most important 
part of evolution that leads to orbital decay, allowing for the formation of close BH-BH binary that 
can merge in the Hubble time. 
}
\label{fig:evol2}
\end{figure}

In Figures \ref{fig:evol1} and \ref{fig:evol2} we present formation scenarios for typical BH-BH 
mergers at redshift $z=6$ in the CE and stable RLOF (standard IMF, delayed SN engine) isolated binary 
evolution channels, respectively. The two chosen examples of BH-BH systems have their total mass and 
metallicity close to the average mass $\overline{M}_{\rm tot,int}=29.3,\ 47.7\msun$ (CE, RLOF) and the 
average metallicity $\overline{Z}=0.0013,\ 0.0012$ (CE, RLOF) at $z=6$ for the two channels (see 
Fig.~\ref{fig:Zz}). For the CE channel we show example with $M_{\rm tot,int}=31\msun$, and for the 
standard RLOF channel with $M_{\rm tot,int}=47\msun$. Note the very clear total mass difference of 
BH-BH mergers from the two formation scenarios. Below we explain the origin of this difference. 

The two examples, despite having rather different primary stars ($\sim38\msun$ and $\sim65\msun$ at 
ZAMS: $t=0$ Myr), produce the similar mass BHs ($\sim14\msun$) out of the primary stars. This comes 
from the fact that both primaries are stripped of their H-rich envelopes (first RLOF phase in each 
case) at different evolutionary stages and their exposed helium cores, Wolf-Rayet (WR) stars, have 
almost the  same mass. In the CE scenario, the first RLOF begins when the primary is on Hertzsprung 
gap and its helium core is developed after the entirety of main-sequence H-burning. In the RLOF 
scenario, the first RLOF begins when the primary is still on the main sequence, and due to mass loss 
the star is de-rejuvenetad, producing a much smaller core than would be expected for a single star 
with the same initial mass. Since both first-formed BHs have the same mass, it means that mass 
difference between the two BH-BH mergers is directly connected to the secondary star, its mass, and 
its evolution. 

In the CE channel, the secondary mass is $34\msun$ at $t=0$ Myr. It increases to $45\msun$ after first 
stable RLOF event. Then the secondary loses mass in stellar winds to initiate the second RLOF at a mass of 
$27\msun$ when it is core He-burning. This second RLOF is a CE event that leaves the secondary as a naked 
helium core (WR star) with a mass of $17\msun$, which quickly collapses to a BH with similar mass. Note 
the significant orbital separation decay during the CE phase, from $a=482\rsun$ to $a=9.5\rsun$. If we made 
the binary at this stage instead proceed through stable RLOF, then there would be not enough orbital 
contraction to produce a merging BH-BH system. Simply, the secondary star would not have enough mass 
in its envelope to allow for enough associated angular momentum loss from the binary to provide 
significant orbital decay. The orbital decay would be rather small: from $a=482\rsun$ to $a=440\rsun$, 
producing a BH-BH system with an inspiral time much larger than the Hubble time. 

In the RLOF channel, the secondary starts with a mass of $64\msun$. It increases to $87\msun$ after the 
first stable RLOF event. Then the secondary, with almost no mass loss: $85\msun$ (mostly main-sequence 
evolution with a relatively small stellar radius), initiates the second RLOF event right after the main 
sequence when it is on the Hertzsprung gap. The second RLOF is stable thermal-timescale mass transfer. 
The RLOF strips the secondary of its envelope, leaving a WR star with a mass of $38\msun$. After losing 
about $5\msun$ to winds, the WR star collapses to a massive BH of $33\msun$. This second RLOF event reduces 
orbital separation from $a=295\rsun$ to $a=6.2\rsun$, producing a close BH-BH system that can merge 
in the Hubble time. The high mass of the donor star (secondary) in this event is required to allow 
for significant decay of the orbit. During thermal-timescale mass transfer, almost the entirety of the 
donor's envelope is lost from the binary (highly over-Eddington mass transfer rate), taking away binary 
angular momentum and reducing the size of the orbit (as long as donor mass is larger than accretor mass). 
In order to form a BH-BH binary system close enough to merge in the Hubble time through two stable RLOF 
episodes, the donor star needs to be few times more massive than its BH companion at the onset of 
the second RLOF event \citep{Heuvel2017,Olejak2021a}. Only such unequal mass binaries 
($q_{\rm RLOF}=\frac{M_{\rm don}}{{M_{\rm acc}}} \gtrsim 3$) are able to eject a large enough amount of mass 
and orbital angular momentum to reduce the orbital separation enough to produce merging BH-BH systems. 

If at onset of the second RLOF event we applied CE evolution instead of stable mass transfer, then the 
ejection of very massive envelope of the secondary ($47\msun$) would lead to post-CE separation of only 
$a=0.7\rsun$. This is smaller than the radius of a secondary core (a WR star: $>1-2\rsun$; 
\cite{Hurley2000}) and thus CE evolution would lead to a merger of the BH with the massive secondary and 
no formation of a BH-BH system. Note that in our standard simulations we would not even attempt to perform 
CE evolution for such a case, as the donor (secondary) is on Hertzsprung gap at the onset of RLOF and 
has a radiative envelope that would not allow for the development/survival of a CE 
\citep{Belczynski2007,Klencki2021}.

Of course, to alleviate this issue one may require the formation of a much wider binary at the onset of 
the second RLOF event in this case. Then, not only may there be enough orbital energy to successfully expel 
the CE, but also the donor star would be core He-burning and much more extended, 
as well as having a convective envelope. However, such star has already evolved through most its life and lost 
a significant fraction of its H-rich envelope through stellar winds. In such a case, even very efficient CE 
evolution may not be enough to decay the orbit below the threshold of formation of merging BH-BH 
systems. It is possible, but it requires fine tuning, and for a given model of CE efficiency and 
adopted stellar winds only few systems form massive BH-BH mergers in the CE channel. This was already 
proposed and explained, although in somewhat different terms, by \cite{Vanson2021}, and our detailed 
description is provided only to serve as a replication study. 

\end{document}